**Elastic Anisotropy Governs the Decay of Cell-induced Displacements**

Shahar Goren[1,#], Yoni Koren[1,#], Xinpeng Xu[2,*], Ayelet Lesman[1,*]

[1] School of Mechanical Engineering, Faculty of Engineering, Tel-Aviv University, Israel

[2] Physics Program, Guangdong Technion—Israel Institute of Technology, Shantou, Guangdong 515063, People's Republic of China

*Correspondence should be addressed to Ayelet Lesman (ayeletlesman@tauex.tau.ac.il) and Xinpeng Xu (xu.xinpeng@gtiit.edu.cn)

#Equal contribution

# Significance

Tissues are made up of cells and an extracellular matrix (ECM), a cross-linked network of stiff biopolymers. Cells actively alter the ECM structure and mechanics by applying contractile forces, which allow them to sense other distant cells and regulate many tissue functions. We study theoretically the decay of cell-induced displacements in fibrous networks, while quantifying the changes in the elastic properties of the cell's local environment. We demonstrate that cell contraction induce an anisotropic elastic state, *i.e.*, unequal principal elastic moduli, in the ECM which dictates the slow decay of displacements. These observations suggest a new mechanical mechanism through which cells can mechanically communicate over long distances, and may provide biomaterials design parameters to guide morphogenesis in tissue engineering.

# Abstract

The unique nonlinear mechanics of the fibrous extracellular matrix (ECM) facilitates long-range cell-cell mechanical communications that would be impossible on linear elastic substrates. Past research has described the contribution of two separated effects on the range of force transmission, including ECM elastic non-linearity and fiber alignment. However, the relation between these different effects is unclear, and how they combine to dictate force transmission range is still elusive. Here, we combine discrete fiber simulations with continuum modeling to study the decay of displacements induced by a contractile cell in fibrous networks. We demonstrate that fiber non-linearity and fiber reorientation both contribute to the strain-induced anisotropy of the elastic moduli of the cell's local environment. This elastic anisotropy is a "lumped" parameter that governs the slow decay of the displacements, and it depends on the magnitude of applied strain, either an external tension or an internal contraction as a model of the cell. Furthermore, we show that accounting for artificially-prescribed elastic anisotropy dictates the displacement decay induced by a contracting cell. Our findings unify previous single effects into a mechanical theory that explains force transmission in fibrous networks. This work provides important insights into biological processes that involve the coordinated action of distant cells mediated by the ECM, such that occur in morphogenesis, wound healing, angiogenesis, and cancer metastasis. It may also provide design parameters for biomaterials to control force transmission between cells, as a way to guide morphogenesis in tissue engineering.

**Introduction**

Cells in tissues are surrounded by a cross-linked network of semi-flexible biopolymers, such as collagen and elastin, known as the extracellular-matrix (ECM) (1, 2). Cells actively adhere and are mechanically connected to the ECM, which enables them to sense and respond to their mechanical microenvironment by applying active contractile forces (3). These cellular forces can alter the structure and mechanical properties of the ECM in the proximity of the cell (4). As such, cell-ECM mechanical interactions play important roles at the cellular level, such as in migration (5, 6), proliferation (6–8), differentiation (5–7, 9, 10), cancer invasion (11–13) and mineral deposition (14).

When cells contract their microenvironment, they cause substantial displacements and structural changes that can reach a distance of tens of cell diameters away (15–18). Such phenomena support long-range cell-cell mechanical communication, a process that can mechanically couple distant cells and coordinate processes such as capillary sprouting (19) and synchronous beating (20). The long-range transmission of cellular forces is usually attributed to the unique nonlinear mechanics and fibrous nature of the ECM (15–17, 21–29). It is currently clear that multiple physical effects (28, 30), including fiber stiffening (15, 31), buckling (17, 22) and collective network responses (*e.g.*, fiber alignment (18, 25, 31)) play central roles in enhancing force transmission in the ECM. Here we show how these effects that were previously considered separately are interlinked and contribute to a more general mechanical mechanism.

Specifically, fibrous networks are known to show substantial stiffening under different kinds of tensile loading (30, 32–35), and contracting cells have been shown to stiffen their surrounding environment when embedded in fibrous gels (15, 34, 36). Stiffening of networks can result either from the stretch-stiffening of individual fibers, as described by the semi-flexible polymer model (2, 37, 38), or from collective network alignment due to strain-induced fiber reorientation (23–25, 39). In various biological contexts, it was frequently shown that fibers align in the vicinity of contracting cells and in the matrix between neighboring cells (12, 13, 18, 23, 26, 27, 31, 36, 40, 41). The tendency of the ECM to stiffen and align under tension was shown to facilitate long-range transmission of internal cellular forces (15, 28, 29, 39, 42). In particular, Hall et al. (16) demonstrated that cell contractility induces local collagen alignment which was shown to be highly correlated to enhanced force transmission. Wang et al. (26) devised a continuum model that incorporates the effect of alignment, and managed to reproduce the nonlinear properties of collagen networks and the increased force transmission. Importantly, however, the isotropic stiffening alone is not enough to account for far-reaching forces observed in experiments, and other mechanisms must be involved (27). For example, buckling of fibers under compression results in softening of fibrous network when compressed over some small critical strain (2, 30, 43–45). In previous works (17, 21, 22, 26, 28) including ours (17, 28), fiber buckling was shown to enhance force transmission. Ronceray et al. (22) has shown, for example, that cell contracting in a simulated fibrous networks creates a buckled region near the cell where the decay of force is very slow. Indeed, fiber buckling has been observed near contracting cells in both experiments (46, 47) and network simulations (22, 48), which provide direct evidence for its involvement. Although the effects of buckling and stiffening have been clearly observed, it is still unknown if they are related phenomena. Taking into account that both buckling (21) and stiffening (15, 31) lead to changes in the elastic properties of the fibrous environment, may hint into the underlying mechanism in controlling force transmission.

Despite the understanding of the effects of fiber nonlinearity (buckling, stiffening) and fiber reorientation to enhancing force transmission, they were previously considered separately. The manner these effects are related and combined to determine the slow decay of cell-induced displacements has yet to be quantified and is still unclear (28, 29). This work combines discrete fiber simulations with continuum modeling to study how the displacements induced by a contracting cell are transmitted in fibrous networks and quantifies how the nonlinear mechanics of individual fibers and collective fiber alignment combine to increase the elastic anisotropy of the cell's local environment. We demonstrate that when a cell pulls on the network, the elastic modulus along the radial direction becomes higher than the modulus along the angular direction, leading to a substantial increase in the elastic anisotropy of the network. This newly established elastic state dictates the decay of displacements in cross-linked fibrous networks. We thus propose a new theory of strain-induced elastic anisotropy through which cells can maintain long-range mechanical communications with one another in fibrous networks consisting of various types of fibers with different properties. We have further validated this mechanism of anisotropy-facilitated displacement transmission by studying the decay of displacements in intrinsically pre-designed anisotropic networks. Our findings provide a new strategy in controlling force transmission in biomaterials by altering the material elastic anisotropy.

## Results and Discussion

**Discrete fiber simulations of a contracting cell in an isotropic network**

Using a simplified finite element model of a contracting cell in a two-dimensional isotropic fibrous network (Fig. 1A), we explore how the nonlinear mechanical properties of the fibers and the strain-induced fiber alignment affect the decay of displacements. The cell is represented by a circular cavity that is isotropically contracting in the fibrous network. The fibrous networks are designed to be *isotropic* (in fiber orientation), and homogenous (in fiber density) at the scale of a cell (network construction and nonlinear properties are described in the Supporting Information, Sec. I and II, respectively). Four fiber models are used to compare different mechanical behaviors of the ECM biopolymers: linear, compression-buckling, tension-stiffening and a model including both buckling and stiffening (Fig. 1B). In all cases, nodes are modeled as freely rotating hinges (*i.e.*, neglecting the rotational resistance of crosslinkers), and only stretching energy is considered. Here we focus on networks with a high connectivity around eight, where the deformations are found to be near-affine (see section III in Supporting Information for simulation evidence of the nearly-affine network deformation). This is in contrast to previous works that concentrated on the classification of different mechanical regimes of fibrous networks, according to connectivity, bending rigidity and internal pre-stresses (39, 49–53). Note that high-connectivity networks can represent fibrous biopolymer gels that are more vigorously cross-linked (54–56) or combined with synthetic gels (57). We take the advantage of the near-affine deformation in such high- connectivity networks to develop and compare with continuum models, which allows us to gain deeper insights into the associated mechanisms. To further simplify the system, we perform the simulations in two-dimensions, which has been shown to qualitatively capture all of the main mechanical behaviors of fibrous ECM, while being substantially simpler computationally (2, 22, 58).

**Two decaying regimes of displacements induced by a contracting cell**

To simulate the decay of network displacements induced by a contracting cell, we embed a cell represented by a circular cavity of radius $R_{\text{cell}}$ in the fibrous network. The cell contracts isotropically by radial displacement $U_{\text{cell}}$ at its edge (Fig. 1A). From these simulations, we obtain the network displacement $U$ as a function of the distance $R$ from the center of the cell (Fig. 2A). We consider cell contractions, $\mathcal{C} \equiv -U_{\text{cell}}/R_{\text{cell}}$, ranging from 2% to 50%. Distances are normalized by $R_{\text{cell}}$ ($\tilde{R} = R/R_{\text{cell}}$) and displacements are normalized by $U_{\text{cell}}$ ($\tilde{U} = U/U_{\text{cell}}$).

Since our network is isotropic at the undeformed state and because all the simulated fiber models behave linearly at small strains, we expect to find $\tilde{U} \sim \tilde{R}^{-1}$ for small strains, as in linear isotropic elastic continuum (59). We indeed observe this scaling at the far-field $\tilde{R} \gg 1$, where strains are small (Fig. 2C). In contrast, close to the cell, up to a distance of around $3R_{\text{cell}}$, the displacement decays more slowly. In this near-field regime, strains are high such that fiber alignment and fiber nonlinearities (stiffening/buckling) play substantial roles in slowing the decay of displacements. We note that the slow decay of displacements in the near-field regime can be well fitted by an effective power law $\tilde{U} \sim \tilde{R}^{-n}$, with $n \leq 1$ for all fiber model types and cell contractions (Fig. 2C). Moreover, we find that as the cell pulls more on the network, the fitted exponent $n$ in the near-field decreases, giving rise to longer range transmission of displacements (Fig. 2D). Remarkably, even for networks of linear fibers, there is a monotonic decrease of $n$ as the cell contracts more. This is in contrast to linear isotropic continuum in which we expect a constant $n = 1$, and the decay of displacement is independent of the cell contraction and the elastic properties of the linear medium (59). The range of displacement transmission is further enhanced (*i.e.*, $n$ decreases more with increasing cell contraction) by introducing fiber buckling and stiffening (inset of Fig. 2D). Interestingly, the exponents $n$ for all types of fibers falls to a single master curve (Fig. 2D) when plotted as a function the *normalized cell contraction*, $\mathcal{C}/\mathcal{C}_{\text{crit}}$, in which $\mathcal{C}_{\text{crit}}$ is the critical cell contraction, a characteristic parameter of the network, at which the near-field region starts to be evident, with *n = 0.9*, as previously defined by Xu et al. (28). This indicates that $\mathcal{C}/\mathcal{C}_{\text{crit}}$, is an essential

"emergent" dimensionless parameter involved in the long-range transmission in fibrous networks; it measures the degree of the nonlinearity in the cell-contracted network.

The increase in the range of displacement transmission by the nonlinear near-field region can be also quantified by introducing an effective cell contraction, $\widetilde{U}_{\text{eff}}$, through fitting the far-field displacement by $\widetilde{U} = \widetilde{U}_{\text{eff}} * \widetilde{R}^{-1}$ (60). $\widetilde{U}_{\text{eff}}$ reflects the degree of cell contraction that would induce the same magnitude of displacements far from the cell in a linear isotropic elastic continuum (Fig. 2E). $\widetilde{U}_{\text{eff}}$ increases with contraction and is enhanced when nonlinear mechanical properties are introduced. We find that for strong cell contraction, $\widetilde{U}_{\text{eff}}$ scales linearly with the normalized cell contraction, $\mathcal{C}/\mathcal{C}_{\text{crit}}$, for all types of fibers simulated in this work. This unified linear relation for all fiber models is consistent with our theory prediction (see Eq. (S36) in Sec. VI in Supporting Information) and further justifies the importance of the normalized cell contraction, $\mathcal{C}/\mathcal{C}_{\text{crit}}$ as an essential "emergent" dimensionless parameter involved in the long-range transmission. Overall, these results indicate that mechanical and geometrical nonlinearities give rise to two power-law regimes with distinct scaling, and nonlinear properties of individual fibers play important roles mainly in the near-field regime. These findings extend on previous studies where multiple scaling regimes were identified (22, 28), by systematically quantifying the effects of the strength of cell contraction and the mechanical properties of fibers on the identified regimes.

**Strain-induced elastic anisotropy: effects of fiber alignment, stiffening and buckling**

Our next goal is to understand why the near-field regime shows a slower decay of displacements, and what mechanism dictates the slope of the decay. We hypothesize that when cells pull on the network, the elastic modulus along the radial direction becomes much higher than the modulus along the angular direction (*i.e.*, the network becomes elastic anisotropic), and that this new elastic state dictates the slow decay of displacements. Our motivation originates from the established theory of anisotropic continuum: when a circular cavity contracts isotropically in a linear anisotropic elastic continuum, the displacements decay as $\widetilde{U} \sim \widetilde{R}^{-\sqrt{E_2/E_1}}$ (21, 61) (Supporting Information, Sec. VI), where $E_1$ and $E_2$ are the stiffness moduli in the radial and angular directions, respectively. Recent experiments by Han et. al. (34) indeed demonstrated that stiffness becomes anisotropic near contracting cells, with a higher extent of stiffening in the radial relative to the angular direction. Moreover, in our simulations, radially aligned fibers are stretched, whereas angularly aligned fibers are typically compressed (Fig. 2B). In the case of nonlinear elastic fibers, this can clearly lead to elastic anisotropy in the network, as radial stretched fibers will become stiffer than the angular compressed fibers.

We first carry out simulations to study how an external strain induces elastic anisotropy in a bulk fibrous network. We generate a rectangular piece of the network and subject it to uniaxial stretch (Fig. 3A, B). We then apply an additional infinitesimal uniaxial stretch to obtain the linear (zero-strain) longitudinal modulus $E_0$. We find that $E_0 \approx \mu \phi_f$ with $\mu$ being the modulus of the fibers and $\phi_f$ being the fiber volume fraction. This agrees with the prediction of classical cellular solid theory (44, 62), justifying our methods of measuring elastic moduli. We then measure the elastic modulus in the transverse direction, and verify that it is equal to the longitudinal value, indicating elastic isotropy of the network at infinitesimal strain (Fig. 3C, for $\epsilon_1 \approx 0$). For each mechanical fiber model, we stretch this bulk network externally over a range of tensile strains $\epsilon_1$, and allow the network to reach equilibrium, leaving free boundary conditions in the transverse direction. The network then assumes an aligned, anisotropic state (Fig. 3B). Upon each external strain $\epsilon_1$, we measure the elastic moduli $E_1$ and $E_2$ along the longitudinal and transverse directions, respectively, by applying an additional infinitesimal tensile stress along each direction (Supporting Information, Section III and Fig. S5).

The *elastic modulus ratio*, $E_2/E_1$, measuring the elastic anisotropy of the network, systematically decreases with increasing external strain (Fig. 3C), showing that the fibrous networks become increasingly anisotropic as they are stretched. Note that this strain-induced anisotropy occurs in networks composed of all model fibers, even for networks composed of *linear* fibers. This indicates that strain-induced elastic anisotropy is driven not only by mechanical nonlinearities of individual fibers, but also by geometrical anisotropy such as collective fiber alignment.

In addition, we find in Fig. 3C that the data points of $E_2/E_1$ for all types of fibers can be approximately fitted by a master curve if plotted versus the normalized strain, $\epsilon/\epsilon_{\text{crit}}$. Here $\epsilon_{\text{crit}}$ is the critical strain, a characteristic parameter of the network, in which $E_2/E_1$ becomes smaller than 0.9, and the strain-induced elastic anisotropy is significant. The normalized strain, $\epsilon/\epsilon_{\text{crit}}$, similar to the normalized cell contraction, $\mathcal{C}/\mathcal{C}_{\text{crit}}$ in cell contraction networks, is another "emergent" dimensionless parameter for the fibrous network upon external uniaxial strain; it measures the degree of the nonlinearity in the externally-stretched network and determines the strain-induced elastic anisotropy.

**Collective fiber alignment measured by nematic order parameter: a characteristic parameter of network elastic anisotropy**

We next aim to quantify the elastic anisotropy that develops in the fibrous network due to cell-induced strains, in the circular geometry as depicted in Fig. 1A. For this purpose, we use fiber alignment as a geometric measure of elastic anisotropy. We assume that if two different networks (both isotropic at undeformed state) are composed of the same type of model fibers and have the same degree of fiber alignment or nematic order parameter (NOP), they should be in the same mechanical state; *i.e.*, they have the same elastic anisotropy. We then approximate the elastic anisotropy of each local area of the network in the cell-contracting system by that of the uniaxially-stretched bulk networks with the same NOP, as measured in Fig. 3C. Thus, we next measure the NOP in both the bulk uniaxial and cell-contraction systems, and use it as a mapping parameter to estimate the elastic anisotropy of the cell system from the measurements of the bulk system.

For rectangular networks under uniaxial strain, we measure the orientation of each fiber with respect to the direction of the applied external strain (Fig. 4A). Before stretching, the network is isotropic and the distribution of fiber angles is uniform without any preferred orientation. After stretching, this distribution shifts towards the direction of the applied stretch (Fig. 4B), indicating a clear collective fiber alignment. The degree of fiber alignment can be measured by the NOP in two dimensions (23): $S = \langle \cos(2\theta) \rangle$. $S$ ranges from $-1$ to $1$, where a value of $0$ corresponds to an isotropic network, a value of $1$ corresponds to a network with fully aligned fibers along the strain direction, and a value of $-1$ corresponds to a network of transversely aligned fibers.

The network comprised of linear fibers shows a linear increase of $S$ with strain, while nonlinear networks exhibit stronger alignment with strain (Fig. 4C, inset). To understand this behavior, we use affine approximation to derive the analytical relation between alignment and external strain, and obtain: $S = \frac{1}{2}(1+\nu)\epsilon_1$, with $\nu = -\epsilon_2/\epsilon_1$ being the Poisson ratio (Supporting Information, Section IV). We find that the theoretical prediction of $S$ is in good agreement with the measured $S$ from the bulk simulations (Fig. 4C), despite possible local non-affine deformations in our fibrous networks (Supporting Information Section III). Thus, a larger Poisson ratio is directly coupled to enhanced alignment. Indeed, nonlinear fiber models are characterized by a larger Poisson ratio that increase with the external strain, whereas linear fiber networks have a relatively constant Poisson ratio ($\nu \sim 0.4$) (Supporting Information, Fig. S6). $S$ therefore depends linearly on the applied strain only for linear networks.

To quantify fiber alignment (*i.e.,* S) in the cell contraction system, we measure the angles relative to the radial direction (Fig. 4D). This is in accordance with the evaluation of alignment described in recent studies (24, 48). $S$ is averaged over fibers in annular domains around the cell. Note that although undeformed networks are macroscopically isotropic, they can locally have some weak preferred fiber orientation. We thus measured the degree of alignment in respect to the undeformed state: $\Delta S = S_{\text{deformed}} - S_{\text{undeformed}}$ for each annular domain around the cell. As shown in Fig. 4E, $\Delta S$ decays with a power-law of $\Delta S \sim R^{-2}$ at distances far from the cell, independent of the degree of cell contraction. This is expected, as $S$ is linearly proportional to strain for small strains (63), and small radial strain in the far-field regimes decays as $1/R^{-2}$ (because strains are the first derivatives of displacements).

In the near-field regime, we observe that $S$ increases linearly with cell contraction and remarkably, it is almost independent of the fiber model type (Fig. 4F), in comparison to the case of uniaxially-stretched networks in rectangular geometry. To understand this better, we derive an analytical prediction for the dependence of $S$ on cell contraction $\mathcal{C}$ in the near-field, using affine displacement approximation with a $\tilde{U} = \tilde{R}^{-n}$ scaling (Supporting Information, Section IV). We obtain the relation: $S = Q(n, \tilde{R}^*)\mathcal{C}$, where $n$ is the near-field power-law exponent and $\tilde{R}^*$ is the size of the near-field region normalized by the cell radius. The exact form of $Q(n, R^*)$ is given in (Supporting Information, Section IV). We find that the dependence of $Q$ on $n$ and $R^*$ is rather weak and $Q(n, R^*)$ is almost constant for all of our simulations. This analytical relation fits nicely to the simulation results (dotted curve in Fig. 4F), which suggests that fiber alignment near the cell can be well predicted from affine theory, and has robust trends independent of fiber mechanics. To conclude, we now have a full description of the network alignment in both the uniaxially-stretched bulk and cell simulation networks.

**Cell-induced elastic anisotropy dictates the slow displacement decay in the fibrous network**

As the cell pulls on the matrix, the network alignment in the near-field increases, and the elastic anisotropy is expected to rise. For each magnitude of cell contraction, we estimate the elastic anisotropy, $1 - \sqrt{E_2/E_1}$, in the near-field based on the measurements in uniaxially-stretched bulk networks composed of the same model fibers and having the same NOP (Fig. 5A, schematics). We can now plot the displacement transmission power-law exponent $n$ against the elastic anisotropy in the near-cell area. We find a linear relation that is similar for all networks types, independent of particular fiber mechanical properties (Fig. 5B). Note that this unified linear relation takes into account the unified master curve for $n$ versus normalized cell contraction $\mathcal{C}/\mathcal{C}_{\text{crit}}$ (Fig. 2D). Furthermore, the obtained unified linear relation indicates that once the elastic anisotropy is set by $\mathcal{C}/\mathcal{C}_{\text{crit}}$, the displacement decay will be dictated accordingly, independent of the fiber mechanical model. Thus, the cell-induced elastic anisotropy is a "lumped" parameter, that takes into account the effects of both fiber nonlinearities and fiber alignment, and which governs the slow decay ($n < 1$) of cell-induced displacements in fibrous networks.

**Controlling the range of displacement transmission by modifying the intrinsic elastic anisotropy**

The change of force transmission by elastic anisotropy can be seen more explicitly by looking at the decay of displacements in an intrinsically anisotropic fiber networks with pre-defined tunable elastic anisotropy. An intrinsically anisotropic fiber network can be constructed by introducing an angle-dependent fiber modulus, for example, $\mu_f = \mu_1 \cos^2\theta + \mu_2 \sin^2\theta$, where $\mu_f$ is the axial fiber modulus, $\theta$ is the fiber orientation (measured relative to the radial direction), $\mu_1$ and $\mu_2$ are two variable stiffnesses (Fig. 6A, inset). Note that in the undeformed state, the network is elastically anisotropic but geometrically isotropic, with a uniform fiber angle distribution. In this case, calculations from affine deformation predicts that the ratio between the two principal moduli takes the form of $E_2/E_1 = (\rho\mu_2 + \mu_1)/(\mu_2 + \rho\mu_1)$, where $\rho = 5.0$ for uniform distribution of fiber orientation [Supporting Information, section V]. We carry out bulk uniaxial simulations to measure $E_2/E_1$ for different predefined ratios $\mu_2/\mu_1$, and obtain $\rho = 4.1$ by least-squares fitting (Fig. 6A). Note that in contrast to the strain-induced anisotropy that is determined by the normalized external strain, $\epsilon/\epsilon_{\text{crit}}$, or normalized internal cell contraction, $\mathcal{C}/\mathcal{C}_{\text{crit}}$, the elastic anisotropy in intrinsically anisotropic networks is determined by design based on predefined fiber anisotropy, $\mu_2/\mu_1$.

We then study the decay of displacements induced by a contractile cell in such intrinsically anisotropic networks. In very small cell contractions, we can consider only the effect of pre-defined elastic anisotropy, while neglecting the effect of fiber alignment. We find that the power-law exponent shows a very good linear proportionality to the elastic anisotropy, $n \approx \sqrt{E_2/E_1}$ for all values of $n$ (Fig. 6B). When the radial direction is stiffer than the angular direction, ($E_2 < E_1$) we obtain $n < 1$, *i.e.,* the displacements decay slowly, in which case, the range of cell-cell communications mediated by the matrix is considerably enhanced. In contrast, when the radial direction is softer than the angular direction ($E_2 > E_1$), we obtain $n > 1$, *i.e.,* the displacement decays *faster* than in linear isotropic elastic medium. In this case, the range of cell-cell communication is restricted. This indicates that the transmission of cellular forces in fibrous networks and hence, the efficiency of matrix-mediated cell-cell communications, can be reprogrammed by modifying the network anisotropic properties.

In summary, we show that the unified mechanism of elastic anisotropy governs the long-range transmission of cell-induced displacement in networks consisting of various types of fibers with different mechanical properties. These findings show that the two effects that were previously considered separately: fiber non-linearity (buckling/stiffening) and fiber reorientation (or alignment), are related and both contribute to the network's elastic anisotropy, which is a key strain-dependent "lamped" parameter that dictates the decay of the displacements. The validity of this unified mechanism of elastic anisotropy is shown in two different cases: cells that contracts strongly enough to generate a local anisotropic environment and cells that contracts weakly in an existing pre-designed anisotropic network. In the former case, the elastic anisotropy is induced by cell contraction and is determined by the normalized cell contraction, $\mathcal{C}/\mathcal{C}_{\text{crit}}$. In the latter case, the elastic anisotropy is intrinsic, pre-designed and is determined by the fiber elastic anisotropy, $\mu_2/\mu_1$. In both cases, the elastic anisotropy dictates the decay of cell-induced displacements.

In this work, we only focused on fibrous networks of high connectivity in two dimensions and neglected the rotational resistance of cross-linkers and the potential feedback of cells to the change in their mechanical microenvironment (3). All these aspects of cell-cell and cell-matrix interactions should be investigated more carefully in the future by extending the continuum theory and finite element simulation presented in this work. Still, this work relates and unifies previous single effects into a general mechanism of elastic anisotropy that explains how (strongly or weakly) contracting cells can maintain long-range mechanical communications with one another. Our findings provide new insights into various biological processes that involve cell-ECM interactions, ECM-mediated cell-cell interactions and for providing a new scaffold design parameter for tissue engineering application.

**Numerical Methods**

We use Matlab to create the network geometry and architecture, and the finite element software Abaqus/CAE 2017 (Dassault Systèmes Simulia) to model the network mechanics and definitions. Truss elements are used to represent the ECM fibers. The response of five networks in different randomnesses is averaged to reduce the effect of the network geometry on the obtained results. The software's implicit static solver is used to simulate the cell contractions for the four fiber models, up to 50%, and the bulk uniaxial and bi-axial stretching for the two fiber models that exhibit strain stiffening. In the case of high tensile strains, we use the implicit dynamic solver for the linear and buckling fiber models, with sufficient damping and loading time that is divided into a large number of increments in order to obtain a quasi-static solution. We ensure that the kinetic energy of the whole model declines to a very small fraction of the internal energy at the end of each stretching (less than 0.001%).

**ACKNOWLEDGMENTS.** We thank Samuel A. Safran from the Weizmann Institute of Science and Yair Shokef from Tel-Aviv University for their useful comments and suggestions. This work was supported by Guangdong Technion–Israel Institute of Technology, by the Israel Science Foundation (1474/16) and Israel Science Foundation- Israeli Centers for Research Excellence (1902/12).

1.  Frantz C, Stewart KM, Weaver VM (2010) The extracellular matrix at a glance. *J Cell Sci* 123(24):4195 LP-4200.
2.  Broedersz CP, MacKintosh FC (2014) Modeling semiflexible polymer networks. *Rev Mod Phys* 86(3):995–1036.
3.  Mohammadi H, McCulloch CA (2014) Impact of elastic and inelastic substrate behaviors on mechanosensation. *Soft Matter* 10(3):408–420.
4.  Jones CAR, et al. (2015) Micromechanics of cellularized biopolymer networks. *Proc Natl Acad Sci* 112(37).
5.  Lo CM, Wang HB, Dembo M, Wang YL (2000) Cell movement is guided by the rigidity of the substrate. *Biophys J* 79(1):144–152.
6.  Wang Y, Wang G, Luo X, Qiu J, Tang C (2012) Substrate stiffness regulates the proliferation, migration, and differentiation of epidermal cells. *Burns* 38(3):414–420.
7.  Xu J, et al. (2017) Effect of matrix stiffness on the proliferation and differentiation of umbilical cord mesenchymal stem cells. *Differentiation* 96:30–39.
8.  Lesman A, Notbohm J, Tirrell DA, Ravichandran G (2014) Contractile forces regulate cell division in three-dimensional environments. *J Cell Biol* 205(2):155–162.
9.  Engler AJ, Sen S, Sweeney HL, Discher DE (2006) Matrix Elasticity Directs Stem Cell Lineage Specification. *Cell* 126(4):677–689.
10. Tony Y, et al. (2004) Effects of substrate stiffness on cell morphology, cytoskeletal structure, and adhesion. *Cell Motil* 60(1):24–34.
11. Ahmadzadeh H, et al. (2017) Modeling the two-way feedback between contractility and matrix realignment reveals a nonlinear mode of cancer cell invasion. *Proc Natl Acad Sci* 114(9).
12. Friedl P, et al. (1997) Migration of Highly Aggressive MV3 Melanoma Cells in 3-Dimensional Collagen Lattices Results in Local Matrix Reorganization and Shedding of α2 and β1 Integrins and CD44. *Cancer Res* 57(10):2061 LP-2070.
13. Shi Q, et al. (2014) Rapid disorganization of mechanically interacting systems of mammary acini. *Proc Natl Acad Sci* 111(2):658 LP-663.
14. Duncan RL, Turner CH (1995) Mechanotransduction and the functional response of bone to mechanical strain. *Calcif Tissue Int* 57(5):344–358.
15. Winer JP, Oake S, Janmey PA (2009) Non-Linear Elasticity of Extracellular Matrices Enables

Contractile Cells to Communicate Local Position and Orientation. *PLoS One* 4(7):e6382.
16. Hall MS, et al. (2016) Fibrous nonlinear elasticity enables positive mechanical feedback between cells and ECMs. *Proc Natl Acad Sci* 113(49):14043–14048.
17. Notbohm J, Lesman A, Rosakis P, Tirrell DA, Ravichandran G (2015) Microbuckling of fibrin provides a mechanism for cell mechanosensing. *J R Soc Interface* 12(108).
18. Harris AK, Stopak D, Wild P (1981) Fibroblast traction as a mechanism for collagen morphogenesis. *Nature* 290:249.
19. Korff T, Augustin HG (1999) Tensional forces in fibrillar extracellular matrices control directional capillary sprouting. *J Cell Sci* 112(19):3249 LP-3258.
20. Nitsan I, Drori S, Lewis YE, Cohen S, Tzlil S (2016) Mechanical communication in cardiac cell synchronized beating. *Nat Phys* 12:472.
21. Rosakis P, Notbohm J, Ravichandran G (2014) A Model for Compression-Weakening Materials and the Elastic Fields due to Contractile Cells. *J Mech Phys Solids* 85:16–32.
22. Ronceray P, Broedersz CP, Lenz M (2016) Fiber networks amplify active stress. *Proc Natl Acad Sci U S A* 113(11):2827–32.
23. Vader D, Kabla A, Weitz DA, Mahadevan L (2009) Strain-Induced Alignment in Collagen Gels. *PLoS One* 4(6):e5902.
24. Abhilash AS, Baker BM, Trappmann B, Chen CS, Shenoy VB (2014) Remodeling of Fibrous Extracellular Matrices by Contractile Cells: Predictions from Discrete Fiber Network Simulations. *Biophys J* 107(8):1829–1840.
25. Aghvami M, Billiar KL, Sander EA (2016) Fiber Network Models Predict Enhanced Cell Mechanosensing on Fibrous Gels. *J Biomech Eng* 138(10).
26. Wang H, Abhilash AS, Chen CS, Wells RG, Shenoy VB (2014) Long-Range Force Transmission in Fibrous Matrices Enabled by Tension-Driven Alignment of Fibers. *Biophys J* 107(11):2592–2603.
27. Rudnicki MS, et al. (2013) Nonlinear Strain Stiffening Is Not Sufficient to Explain How Far Cells Can Feel on Fibrous Protein Gels. *Biophys J* 105(1):11–20.
28. Xu X, Safran SA (2015) Nonlinearities of biopolymer gels increase the range of force transmission. *Phys Rev E* 92(3):32728.
29. Zhang Y, Feng J, Heizler SI, Levine H (2018) Hindrances to precise recovery of cellular forces in fibrous biopolymer networks. *Phys Biol* 15(2):26001.
30. Van Oosten ASG, et al. (2016) Uncoupling shear and uniaxial elastic moduli of semiflexible biopolymer networks: Compression-softening and stretch-stiffening. *Sci Rep* 6.
31. Ma X, et al. (2013) Fibers in the Extracellular Matrix Enable Long-Range Stress Transmission between Cells. *Biophys J* 104(7):1410–1418.
32. Storm C, Pastore JJ, MacKintosh FC, Lubensky TC, Janmey PA (2005) Nonlinear elasticity in biological gels. *Nature* 435:191.
33. Wen Q, Basu A, Winer JP, Yodh A, A Janmey P (2007) Local and global deformations in a strain-stiffening fibrin gel. *New J Phys* 9(11):428.
34. Han YL, et al. (2017) Cell contraction induces long-ranged stress stiffening in the extracellular matrix. *Proc Natl Acad Sci* 115(16):4075 LP-4080.
35. Kang H, et al. (2009) Non-linear elasticity of stiff filament networks: Strain stiffening, negative normal stress, and filament alignment in fibrin gels. *J Phys Chem B* 113(12):3799–3805.
36. Jansen KA, Bacabac RG, Piechocka IK, Koenderink G (2013) Cells Actively Stiffen Fibrin Networks by Generating Contractile Stress. *Biophys J* 105(10):2240–2251.
37. Žagar G, Onck PR, Van Der Giessen E (2015) Two fundamental mechanisms govern the

stiffening of cross-linked networks. *Biophys J* 108(6).
38. Heidemann KM (2016) On the mechanics of biopolymer networks. Dissertation (der Georg-August University School of Science (GAUSS)).
39. Feng J, Levine H, Mao X, Sander LM (2016) Nonlinear elasticity of disordered fiber networks. *Soft Matter* 12(5).
40. Fraley SI, et al. (2015) Three-dimensional matrix fiber alignment modulates cell migration and MT1-MMP utility by spatially and temporally directing protrusions. *Sci Rep* 5.
41. Sopher RS, et al. (2018) Nonlinear Elasticity of the ECM Fibers Facilitates Efficient Intercellular Communication. *Biophys J* 115(7):1357–1370.
42. Feng J, Levine H, Mao X, Sander LM (2015) Alignment and nonlinear elasticity in biopolymer gels. *Phys Rev E - Stat Nonlinear, Soft Matter Phys* 91(4).
43. Kim O V, et al. (2016) Foam-like compression behavior of fibrin networks. *Biomech Model Mechanobiol* 15(1):213–228.
44. Xu X, Safran SA (2017) Compressive elasticity of polydisperse biopolymer gels. *Phys Rev E* 95(5):52415.
45. Blundell JR, Terentjev EM (2009) Buckling of semiflexible filaments under compression. *Soft Matter* 5(20):4015–4020.
46. Burkel B, Notbohm J (2017) Mechanical response of collagen networks to nonuniform microscale loads. *Soft Matter* 13(34):5749–5758.
47. Liang L, Jones C, Chen S, Sun B, Jiao Y (2016) Heterogeneous force network in 3D cellularized collagen networks. *Phys Biol* 13(6)
48. Humphries DL, Grogan JA, Gaffney EA (2017) Mechanical Cell–Cell Communication in Fibrous Networks: The Importance of Network Geometry. *Bull Math Biol* 79(3):498–524.
49. Alvarado J, Sheinman M, Sharma A, MacKintosh FC, Koenderink G (2017) Force percolation of contractile active gels. *Soft Matter* 13(34).
50. Broedersz CP, Mao X, Lubensky TC, MacKintosh FC (2011) Criticality and isostaticity in fibre networks. *Nat Phys* 7:983.
51. Broedersz CP, Sheinman M, MacKintosh FC (2012) Filament-Length-Controlled Elasticity in 3D Fiber Networks. *Phys Rev Lett* 108(7):078102.
52. Piechocka IK, Bacabac RG, Potters M, MacKintosh FC, Koenderink G (2010) Structural hierarchy governs fibrin gel mechanics. *Biophys J* 98(10):2281–2289.
53. Head DA, Levine AJ, MacKintosh FC (2003) Distinct regimes of elastic response and deformation modes of cross-linked cytoskeletal and semiflexible polymer networks. *Phys Rev E* 68(6):61907.
54. Davidenko N, et al. (2015) Control of crosslinking for tailoring collagen-based scaffolds stability and mechanics. *Acta Biomater* 25:131–142.
55. Berkache K, Deogekar S, Goda I, Catalin Picu R, Ganghoffer J-F (2018) Identification of equivalent couple-stress continuum models for planar random fibrous media. *Contin Mech Thermodyn*.
56. Khalily MA, Goktas M, Guler MO (2015) Tuning viscoelastic properties of supramolecular peptide gels via dynamic covalent crosslinking. *Org Biomol Chem* 13(7):1983–1987.
57. Burla F, et al. (2019) Stress management in composite biopolymer networks. *Nat Phys*.
58. Head DA, Levine AJ, MacKintosh FC (2003) Deformation of Cross-Linked Semiflexible Polymer Networks. *Phys Rev Lett* 91(10):108102.
59. Landau 1908-1968 LD (Lev D (1959) *Theory of elasticity / by L. D. Landau and E. M. Lifshitz ; translated from the Russian by J. B. Sykes and W. H. Reid* ed Lifshitz EM (Evgenii M (Pergamon,

London). linear ela.
60. Shokef Y, Safran SA (2012) Scaling Laws for the Response of Nonlinear Elastic Media with Implications for Cell Mechanics. *Phys Rev Lett* 108(17):178103.
61. Lekhtnitskii SG (1981) *Theory Of Elasticity Of An Anisotropic Body*.
62. J. A. A. Gibson L, F. Ashby M (1988) *Cellular Solids: Structure And Properties*
63. Fischer-Friedrich E (2018) Active Prestress Leads to an Apparent Stiffening of Cells through Geometrical Effects. *Biophys J* 114(2):419–424.

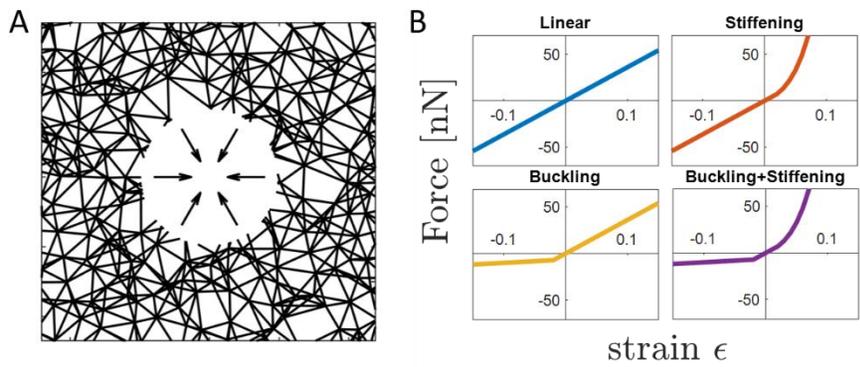

*Figure 1: Cell contraction simulation. (A) Illustration of a contractile cell, simulated by a circular cavity, embedded in an isotropic, homogeneous fibrous network. Isotropic inward displacements are imposed at the boundary of the cell, and the outer boundary of the network is held to be fixed (not shown). (B) Force-strain curves for individual fibers used in the simulations. Four different mechanical models are used: linear, stretch-stiffening, compression-buckling, and nonlinear fibers with both buckling and stiffening.*

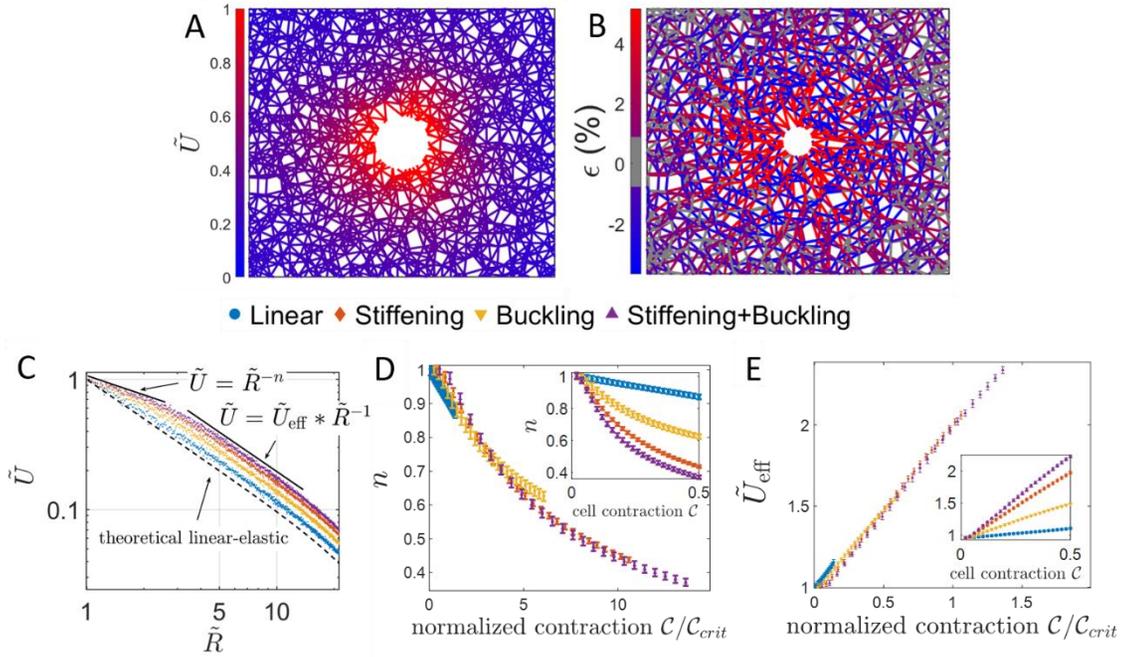

*Figure 2: Decay of displacements induced by isotropic cell contraction. (A) Color map of displacements induced by 50% cell contraction. (B) Color map of strains of individual fibers. Radially aligned fibers are stretched (red) while angularly aligned fibers are compressed (blue). (C) Normalized displacement $\tilde{U}$ as a function of the normalized distance $\tilde{R}$ from the cell center, for the four fiber models, with 40% cell contraction. Two power-law regimes can be identified. (D) Near-field effective power-law exponent n, versus normalized cell contraction, $\mathcal{C}/\mathcal{C}_{crit}$. Inset: n decreases linearly with contraction for linear fibers, due to increased fiber alignment. Fiber nonlinearities facilitate the decrease in n. Error bars are averaged over 5 network realizations. (E) Effective cell displacement $\tilde{U}_{eff}$ versus normalized cell contraction, $\mathcal{C}/\mathcal{C}_{crit}$. $\tilde{U}_{eff}$ increased linearly with contraction for all fiber models (inset).*

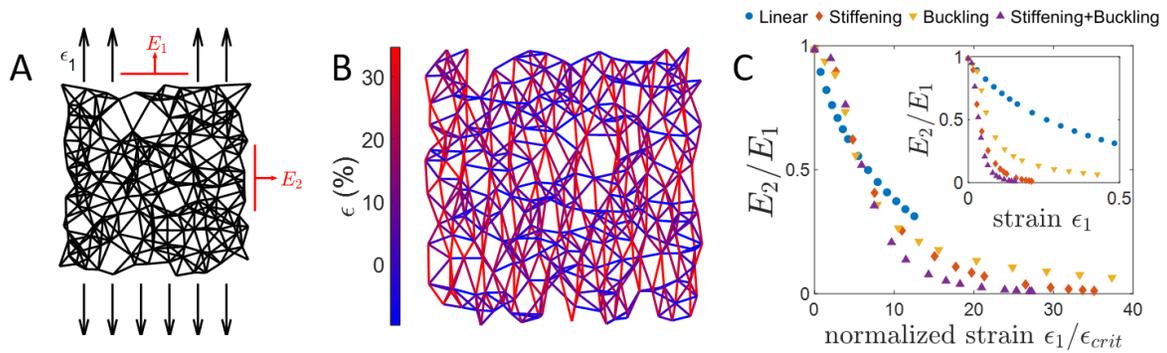

*Figure 3: Elastic anisotropy induced by external uniaxial tension. (A) Schematic illustration of the rectangular network upon external uniaxial stress. (B) Strain color-map: The fibers along the axis of strain are stretched (red) while transverse fibers are compressed (blue). (C) The ratio of the two principal moduli, $E_2/E_1$, measuring the elastic anisotropy, versus normalized cell contraction, $\epsilon_1/\epsilon_{crit}$. At a given tensile pre-strain $\epsilon_1$, additional infinitesimal tensile strains are applied in both directions to measure the respective elastic moduli, $E_1$ (longitudinal) and $E_2$ (transverse) (see illustration in a). The modulus ratio, $E_2/E_1$, deviates from 1 when the network is strained, and becomes close to zero for high strains. All the three effects – fiber alignment, buckling and stiffening, contribute to this strain-induced anisotropy. Inset shows $E_2/E_1$ versus cell contraction (without normalization).*

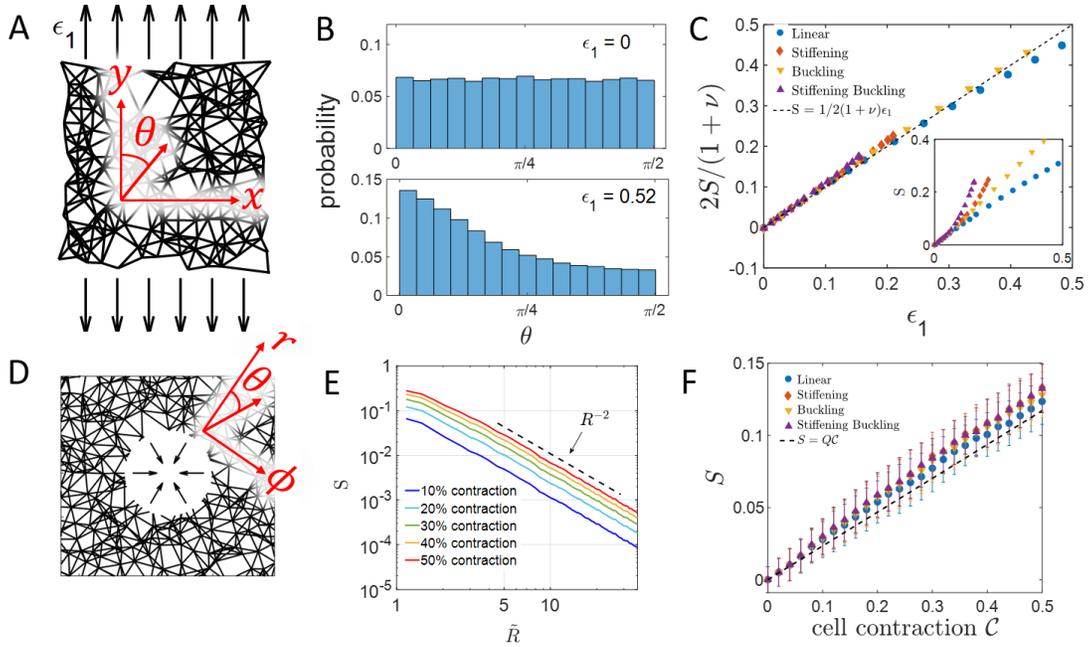

*Figure 4: Strain-induced collective alignment of fibers. (A) Schematic illustration of uniaxially-stretched network. The orientational angle, θ, of each fiber is defined with respect to the strain axis. (B) Histogram of the distribution of individual fiber angles. (top) Fiber angle distribution in an undeformed isotropic network. (bottom) Fiber angle distribution for a linear fiber network upon 0.52 uniaxial strain; collective fiber alignment along the strain direction is indicated with more fiber fraction towards zero angle. (C) Nematic order parameter S of the stretched network as a function of the applied uniaxial strain $\epsilon_1$. All fiber models follow the affine prediction $S = 1/2(1 + \nu)\epsilon_1$. (D) Schematic illustration of a cell-contracted network. The orientational angle, θ, of each fiber is defined with respect to the radial direction. (E) Nematic order parameter S of the contracted network (of linear fibers) as a function of $\tilde{R}$. S decays with $\sim R^{-2}$ in the far field with a slower decay in the vicinity of the cell. S increases with cell contraction. (F) Averaged S in the near-field region as a function of cell contraction. S increases linearly with cell contraction, and fits well to the affine theory (dashed black line). Error bars were calculated based on the average of 5 network realizations.*

Page **19** of **32**

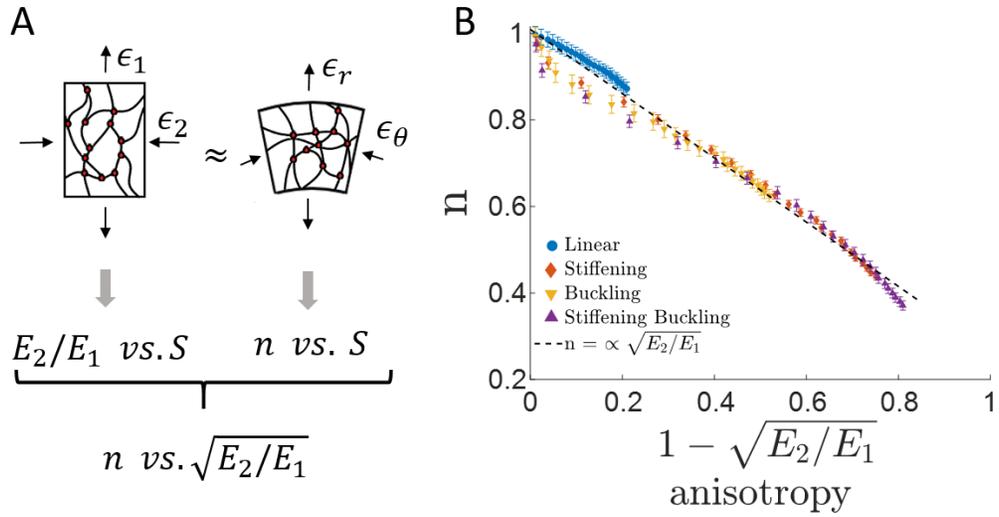

*Figure 5: Elastic anisotropy dictates the decay of displacements in fibrous networks. A Schematic illustration of the transformation from bulk to cell contraction simulations: If two different networks (both isotropic at undeformed state) are composed of similar fibers have the same nematic order parameter, they are assumed to be in the same mechanical state, i.e., have the same elastic anisotropy, $E_2/E_1$. (B) The near-field power-law exponent n for the decay of the displacement is plotted as a function of the square root of the network anisotropy $\sqrt{E_2/E_1}$. We found $n \propto \sqrt{E_2/E_1}$ which is consistent with the theoretical prediction for an anisotropic continuum material.*

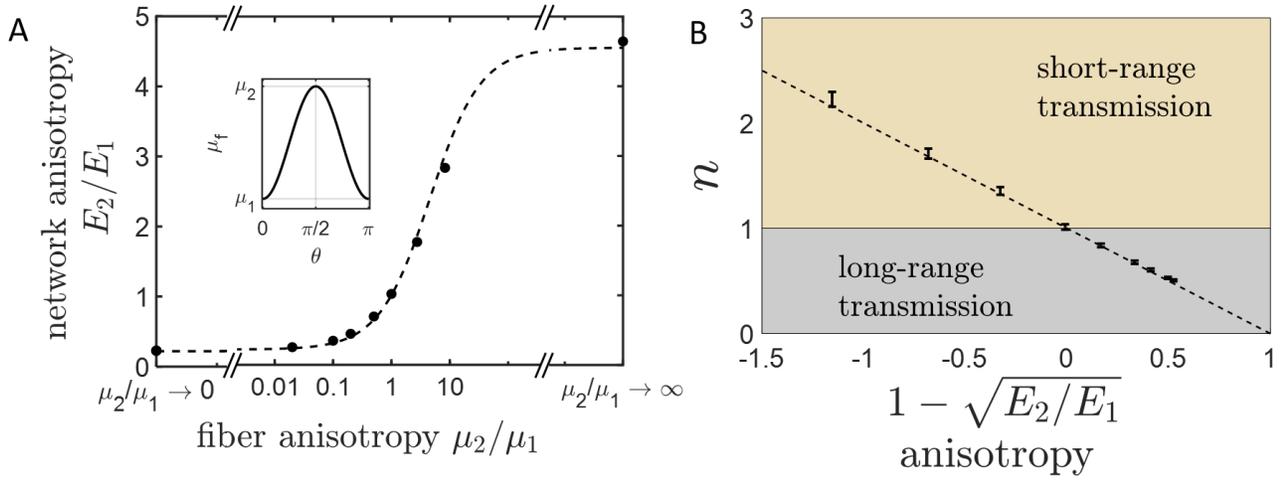

*Figure 6: The decay of cell-induced displacements in intrinsically anisotropic networks. (A) An intrinsically anisotropic network is constructed by introducing an orientation-dependent fiber modulus $\mu_f = \mu_1 \cos^2(\theta) + \mu_2 \sin^2(\theta)$ (inset). The elastic anisotropy of the network is predicted by affine theory, as $E_2/E_1 = (\rho\mu_2 + \mu_1)/(\rho\mu_1 + \mu_2)$ (dashed line) with $\rho \approx 4.1$ (black dots, simulation data) which are closed to the theory prediction $\rho = 5.0$. (B) The near-field power-law exponent, n, is plotted as a function of network anisotropy at infinitesimal cell contractions; a very good linear fitting to $n = \sqrt{E_2/E_1}$ (dashed line) is obtained in both regions: $E_2 > E_1$ (with n>1 indicating fast displacement decay) and $E_1 > E_2$ (with n<1 indicating slow displacement decay).*

**Supporting Information**

**In** this supporting information, we provide all the details about the method of discrete fiber simulations of fibrous networks, quantification of elastic anisotropy and fiber alignment, and the simplified continuum model we are using.

**I. Mechanical properties of four model fibers**

    Each individual fiber is modeled as a one-dimensional linear truss element undergoing uniaxial tension or compression. The four material models used to simulate the mechanical elastic behavior of the individual fibers ( Fig. 1A in the main text) are modeled as follows: linear fibers have an elastic modulus of $11.5 kPa$, both tensile and compressive. For buckling fibers, the elastic modulus is ten times smaller in compressive strains larger than 2%. Stiffening fibers in tension are characterized by elastic modulus, which increases exponentially in tensile strains above 2% and is constant in lower strains. Buckling-stiffening fibers combine both softening and stiffening of the two last mentioned material models,

respectively. The elastic modulus assigned to each of the individual fibers exhibit buckling and stiffening is given by

$$E = \begin{cases} \rho E^*, & \varepsilon < \varepsilon_b \\ E^*, & \varepsilon_b < \varepsilon < \varepsilon_s \\ E^* \exp[(\varepsilon - \varepsilon_s)/\varepsilon_0], & \varepsilon > \varepsilon_s \end{cases} \qquad (S1)$$

while the values used in the model are $E^* = 11.5 kPa$ as the given elastic modulus, $\varepsilon_b = 2\%$ is the minimal compressive strain for which buckling occurs, $\varepsilon_s = 2\%$ is the tensile strain above which strain stiffening occurs, $\varepsilon_0 = 5\%$ is the strain-stiffening coefficient and $\rho = 0.1$ is the buckling ratio, which is the ratio between the compressive and tensile elastic moduli. This mechanical behavior is largely based on computational models described in previous studies (1).

## II. Numerical construction of an isotropic and homogeneous fibrous network

First, we randomly insert some nodes in two dimensions throughout a circular domain of radius $R_b$ by following a uniform distribution. Pairs of nodes are then connected by single fibers generated following a minimum cost algorithm as follows. The probability that a potential fiber connecting two neighboring nodes would be generated is determined by a cost function, $P$. In every iteration, each node looks at the thirty nearest neighbors in its vicinity as potential fibers to be generated. The cost function associated with each potential fiber is given by:

$$P = N_{ij} + aC_{ij} + b(A_i + A_j) \qquad (S2)$$

with the two constants $a$ and $b$ determining the relative importance of each term. A potential fiber can be generated only if its cost function, $P$, is negative. The three terms in Eq. (S2) are explained as follows.

(i) $N_{ij}$, is the degree of nearness of each node $i$ to the other node $j$. For example, if node $i$ is the second nearest neighbor of node $j$, and node $j$ is the third nearest neighbor of node $i$, then $N_{ij} = 2 + 3 = 5$.

(ii) $C_{ij} = c_i + c_j - 2c_{opt}$ takes into account of the current connectivity numbers $c_i$ and $c_j$ of nodes $i$ and $j$, respectively, with $c_{opt}$ being a chosen optimal connectivity. We can set the average connectivity of the network by tuning $c_{opt}$. The higher $c_{opt}$ is, the smaller $C_{ij}$ becomes and so the probability for creating the fiber gets higher.

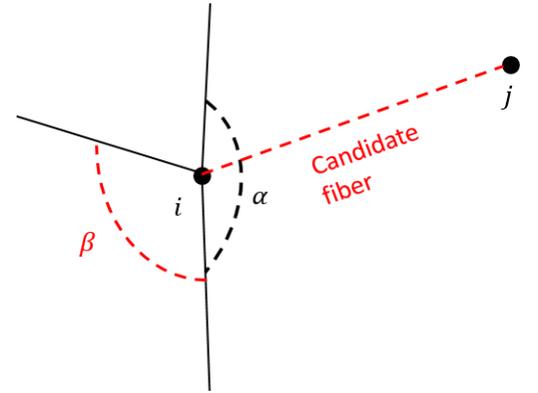

Fig. S1: Schematic illustration of local geometrical consideration in generating new fibers. For a new candidate fiber between nodes, $i$, and $j$, we define $A_i = \beta_i - \alpha_i$, where $\alpha_i$ and $\beta_i$ are the maximum free angles before and after adding the new fiber. If $A_i$ is negative, the fiber will have a lower cost. The same is done for node $j$.

(iii) The last term relates to the maximum free angle of each node (as shown in Fig. S1), representing the local geometry around the nodes. Here $A_i$ is the difference in maximum free angle after and before adding the new potential fiber to node $i$. A negative $A_i$ increases the probability that a fiber would be generated.

In summary, in every iteration, each node is connected by a new fiber to the node with the minimal (negative) cost (in comparison to the other twenty-nine nearest neighbors). The iteration process finishes when no new fiber can be added.

The major advantages of this method of network construction are as follows. (i) The generated networks are therefore always isotropic. Nodes tend to be connected by fibers that are oriented in a wide range of directions, in comparison to other methods where many fibers that are aligned mostly in some particular directions. This increases the isotropy and homogeneity of the networks. (ii) It allows for control of geometrical features like pore size and connectivity in an elegant way. (iii) It does not depend on macroscopic length scales such as cell and system sizes, but only depends on the density of nodes.

In addition, for the simulations of networks that are contracted by an embedded cell, we remove all fibers or fiber sections inside the cell area, not affecting the fibers outside. The cell edge remains linked with the surrounding network. This guarantees that the procedure of embedding the cell in the network does not change the geometry around it. For bulk simulations with a network stretched by uniaxial stresses, a rectangular section of network is chosen and the rest of the network is removed.

### III. Nonlinear elastic responses of fibrous networks to external uniaxial stretch

*1. Nearly-affine deformations of the network*

Our networks are classified as high-connectivity, stretch dominated networks, and are therefore expected to deform in an affine manner. For affine deformations, the coordinated of the nodes are transformed by $x \to x(1 + \epsilon_1); \; y \to y(1 + \epsilon_2)$, where $\epsilon_1, \epsilon_2$ are the macroscopic principal strains. From this it can be proved that the strain and orientation of each fiber following deformation, depend only on its initial orientation and the macroscopic network principal strains, and are given by:

$$\epsilon_f = \sqrt{(1 + \epsilon_1)^2 \cos^2(\theta_0) + (1 + \epsilon_2)^2 \sin^2\theta_0} - 1 \quad (S3)$$

$$\cos^2(\theta) = \frac{(1 + \epsilon_1)^2 \cos^2(\theta_0)}{(1 + \epsilon_1)^2 \cos^2(\theta_0) + (1 + \epsilon_2)^2 \sin^2(\theta_0)} \quad (S4)$$

where $\epsilon_f$ and $\theta$ are the strain and final orientation of the fiber, $\theta_0$ is the initial fiber orientation.

The following figures compare our simulation results to these predictions. The figures are shown only for linear fibers and stiffening-buckling fibers.

Figures S2-S4 indicate that our network deforms in a way that is close to affine over all strains and fiber models. Large strains and fiber nonlinearity increase the local deviations from the affine predictions, but the overall affine trend is roughly preserved. Notice that in general, the average strains in the network are smaller than the affine predictions, leading to lower deformation energies. Notice also the large Poisson effect for nonlinear fibers.

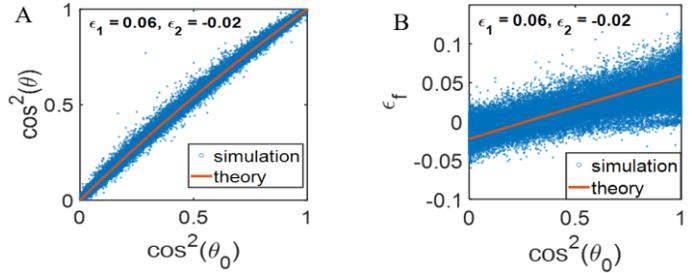

*Fig. S2: Comparison of affine predictions (Eqs. (S3) and (S4)) to simulations, for <u>Linear</u> fibers under small strains.*

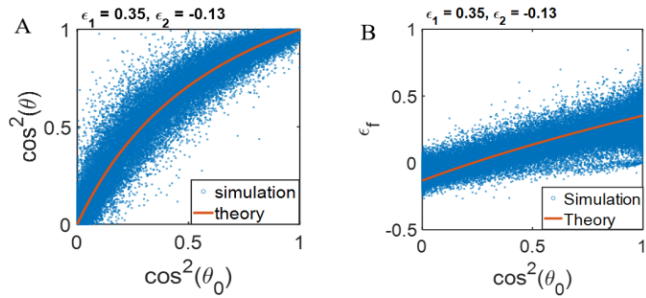
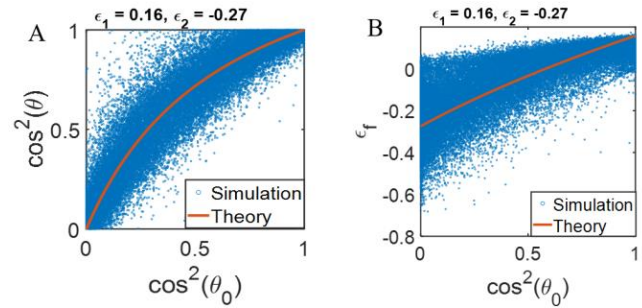

*Fig. S3: Comparison of affine predictions (Eqs. (S3) and (S4)) to simulations, for <u>Linear</u> fibers under large strains.*

*Fig. S4: Comparison of affine predictions (Eqs. (S3) and (S4)) to simulations, for <u>Stiffening-Buckling</u> fibers under large strains.*

*2. Stretch-induced anisotropy and Poisson ratio of the network*

The initially (before deformation) isotropic and homogeneous fibrous networks show highly nonlinear elastic responses to the applied external uniaxial stretch. On the one hand, the Poisson ratio, $\nu_0$, depends significantly on the applied uniaxial strain, $\epsilon_1$ (See Fig. S5), except for the networks of linear fibers where $\nu_0 \approx 0.4$. $\nu_0$ increases from its linear value ~0.4 to be even larger than 1.6, beyond the upper limit, 1.0, of Poisson ratio in 2D linear isotropic, elastic continuum. Note that the volume change is described by strain $\epsilon_1 + \epsilon_2 = \epsilon_1(1 - \nu_0)$; that is the network volume increases if $\nu_0$<1 and decreases if Poisson ratio is greater than 1 for the networks of stiffening fiber, i.e., $\nu_0$>1. This indicates the presence of nonlinearity as well as fiber dilution/ densification in the deformed fibrous networks.

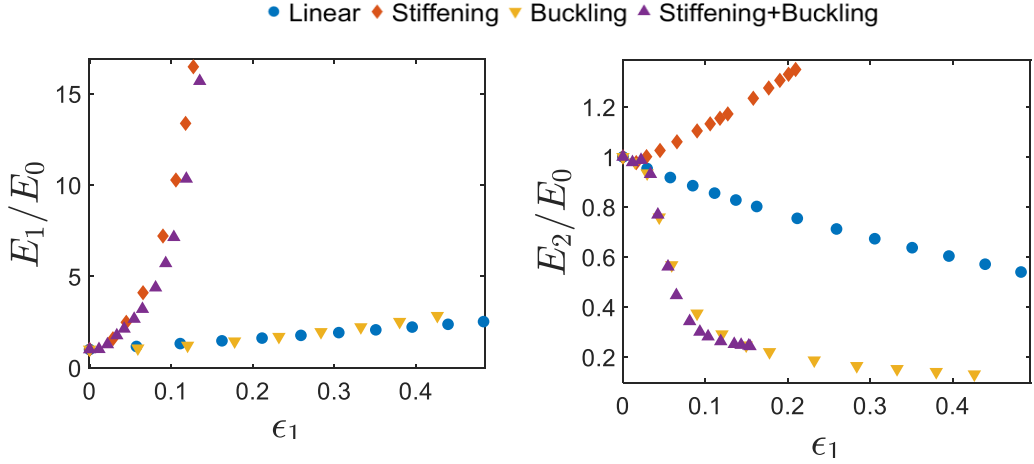

Fig. S5. Two principal moduli, $E_1$ and $E_2$, along the longitudinal and transverse directions, respectively, as a function of the applied strain, $\epsilon_1$. Elastic anisotropy, represented by the inequality of $E_1$ and $E_2$, is shown to be induced by the applied external uniaxial strain.

On the other hand, the initially isotropic network will become anisotropic in elasticity when it is stretched strongly enough. Fig. S6 shows that the two principal moduli, $E_1$ and $E_2$, become unequal over some threshold strain, representing elasticity anisotropy induced by the applied strain, $\epsilon_1$. Note that the longitudinal modulus, $E_1$, increases with strain in all the cases. The curve of linear fibers overlaps that of buckling fibers, while the curve of stiffening fibers overlaps that of stiffening-buckling fibers; the slope of later curve is much larger than the former one due to stretch-stiffening nonlinearity. In comparison, the transverse modulus, $E_2$, decreases with strain in all the cases except for stiffening fibers. The curve of buckling fibers overlaps that of stiffening-buckling fibers, the slope magnitude of which is much larger than that of linear fibers due to the compression-buckling nonlinearity in the transverse directions. Interestingly, the transverse modulus, $E_2$, of stiffening fibers increases (not decreases as the other fiber types) with strain, $\epsilon_1$. However, when we plot $E_2/E_1$ as a function of the applied strain, $\epsilon_1$ (as shown in Fig. 3 in the main text), we find $E_2/E_1$ is always smaller than 1.0 for all the fiber types, although the magnitudes of $E_2/E_1$ for given strain depend significantly on the particular fiber type.

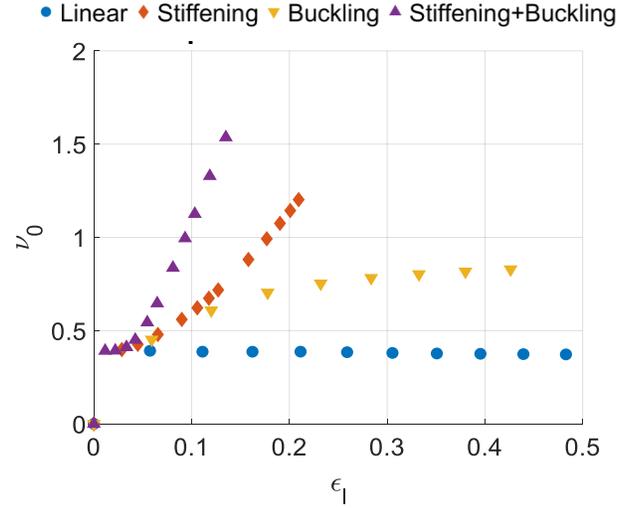

Fig. S6: Poisson ratio, $\nu_0$, as a function of applied uniaxial strain, $\epsilon_1$ for fibrous networks composed of four different model fibers.

### IV. Strain-induced collective fiber alignment

*1. Fiber alignment induced by external uniaxial stretch*

The external uniaxial stress applied on the fibrous networks will induce anisotropy not only in elasticity but also in geometry -- nematic order accompanying collective fiber alignment. To quantify the collective uniaxial fiber alignment in 2D networks upon external uniaxial stretch, we employ the 2D nematic order parameter:

$$S \equiv \langle \cos(2\theta) \rangle = \int p(\theta) \cos(2\theta)\, d\theta \qquad (S5)$$

where $\theta \in [0, \pi)$ is the fiber orientation with respect to the direction of the applied stretch and $p(\theta)$ is the probability distribution function of the fiber orientation.

For initially (before stretched) isotropic networks, we assume the fiber angle distribution to be uniform, i.e., $p(\theta) = 1/\pi$, and the corresponding nematic order parameter is $S_0 = 0$. For deformed fibrous networks with two principal strains, $\epsilon_1$ and $\epsilon_2$, if we assume affine deformations of each fibers, we then obtain the new angle, $\theta$, of the fiber with initial angle, $\theta_0$, is given by

$$\theta = \mathrm{atan}\left(\tan(\theta_0) \frac{1 + \epsilon_1}{1 + \epsilon_2}\right), \qquad (S6)$$

which in the case of small strains ($\epsilon_1, \epsilon_2 \ll 1$) can be expanded to the first order in strains as

$$\theta \approx \theta_0 + \frac{\epsilon_1 - \epsilon_2}{2} \sin(2\theta_0). \qquad (S7)$$

Moreover, we can get the probability distribution of fiber orientation in deformed networks as

$$p(\theta) = p(\theta_0) \frac{d\theta_0}{d\theta} \approx \frac{1}{\pi}[1 + (\epsilon_1 - \epsilon_2)\cos(2\theta)], \qquad (S8)$$

from which we obtain

$$S = \frac{1}{\pi} \int (\epsilon_1 - \epsilon_2) \cos^2(2\theta)\, d\theta = \frac{\epsilon_1 - \epsilon_2}{2}. \qquad (S9)$$

Particularly, in the case of uniaxial deformation with $\epsilon_2 = -\nu\epsilon_1$, we obtain

$$S = \frac{(1 + \nu)}{2} \epsilon_1. \qquad (S10)$$

Note that Poisson ratio is usually not a constant but a nonlinear function of strain $\epsilon_1$ as shown in Fig. S5. Therefore, the nematic order parameter induced by strain is generally a nonlinear function of strain as shown in Fig. 4C (in the main text) and is linearly proportional to strain only if Poisson ratio is a constant (also see Fig. 4C). We have compared the theoretical prediction Eq. (S10) with simulation results; very good agreement is achieved as shown in Fig. 4C. There, we plot $2S/(1+\nu)$ as a function of $\epsilon_1$ and find that all the data obtained from four types of fiber networks fall onto one single master curve as indicated by Eq. (S10).

*2. Fiber alignment induced by internal cell contraction*

Let's now consider the collective fiber alignment in fibrous networks that is induced by internal cell contraction. In the near-field region, the decay of the cell-induced displacement follows a power law

$$u = U_{\text{cell}} \left(\frac{R}{R_{\text{cell}}}\right)^{-n}, \qquad (S11)$$

with $R_{\text{cell}}$ and $U_{\text{cell}} < 0$ being the cell radius and its contracting displacement, respectively, from which we obtain the two principal strains as

$$\epsilon_1 = \frac{\partial u}{\partial R} = -n U_{\text{cell}} \frac{R^{-n-1}}{R_{\text{cell}}^{-n}}, \quad \epsilon_2 = \frac{u}{R} = U_{\text{cell}} \frac{R^{-n-1}}{R_{\text{cell}}^{-n}}. \qquad (S12)$$

In this case, we obtain from Eq. (S9) that the nematic order parameter, $S$, of small section of the network at $R$, is given by

$$S(R) = -\frac{1+n}{2} U_{\text{cell}} \frac{R^{-n-1}}{R_{\text{cell}}^{-n}}. \qquad (S13)$$

Averaging $S$ over the near-field region from the cell boundary $R = R_{\text{cell}}$ to some characteristic length $R = R^*$, we obtain the averaged near-field $S$ as

$$\langle S \rangle = \frac{\int_{R_{\text{cell}}}^{R^*} S(R) R dR}{\int_{R_{\text{cell}}}^{R^*} R \, dR} = Q(n, R^*) \frac{U_{\text{cell}}}{R_{\text{cell}}}, \qquad (S14)$$

with

$$Q(n, R^*) \equiv -\frac{1+n}{1-n} \frac{\tilde{R}^{*1-n} - 1}{\tilde{R}^{*2} - 1}, \qquad (S15)$$

and $\tilde{R}^* \equiv R^*/R_{\text{cell}}$. We have also compared the theoretical prediction (S14) with simulation results. All the data obtained from four different types of fiber networks fall on one curve which agrees with (S14) very well as shown in Fig. 4F in the main text.

## V. Anisotropic elasticity of an intrinsically anisotropic network

We construct a homogeneous fibrous network composed of linear fibers with intrinsic anisotropy by following the same method as explained in Sec. I with uniform fiber angle distribution, i.e., $p(\theta) = 1/\pi$ for $\theta \in [0, \pi)$, and we introduce an orientation-dependent stiffness, $\mu_f$, to each fiber as

$$\mu_f = \mu_1 \cos^2(\theta) + \mu_2 \sin^2(\theta), \qquad (S16)$$

where $\theta$ is the angle of the fiber with respect to radial direction of the cell, $\mu_1$ and $\mu_2$ being the two extrema of $\mu_f$ along the radial (with $\theta = 0$) and angular (with $\theta = \pi/2$) directions, respectively. That is, we have constructed a fibrous network with intrinsic anisotropy in elasticity but not in geometry (without collective fiber alignment).

If the fibers in the network are deformed affinely, the strain of a fiber with orientation $\theta$ is:

$$\epsilon_f = \sqrt{(1+\epsilon_1)^2 \cos^2(\theta) + (1+\epsilon_2)^2 \sin^2(\theta)} - 1, \qquad (S17)$$

where $\epsilon_1, \epsilon_2$ are the two principal strains along the radial and angular directions, respectively. Then the corresponding deformation energy of the fiber is

$$\begin{aligned}F_f(\theta, \epsilon_1, \epsilon_2) &= \frac{l^2}{2}\mu_f \epsilon_f^2 \\ &= \frac{l^2}{2}(\mu_1 \cos^2(\theta) + \mu_2 \sin^2(\theta))\left(\sqrt{(1+\epsilon_1)^2 \cos^2(\theta) + (1+\epsilon_2)^2 \sin^2(\theta)} - 1\right)^2, \end{aligned} \qquad (S18)$$

with $l$ being the undeformed length of the fiber. For small strains, $F_f(\theta, \epsilon_1, \epsilon_2)$ can be expanded in $\epsilon_1, \epsilon_2$ to the second order as

$$\begin{aligned}F_f \approx \frac{l^2}{2}(\mu_1 \cos^2\theta + \mu_2 \sin^2\theta)&\left[1 + (1+\epsilon_1)^2 \cos^2\theta + (1+\epsilon_2)^2 \sin^2\theta \right.\\ &\left. - 2\left(1 + \epsilon_1 \cos^2\theta + \epsilon_2 \sin^2\theta + \frac{\epsilon_1^2}{2}(\cos^2\theta - \cos^4\theta) + \frac{\epsilon_2^2}{2}(\sin^2\theta - \sin^4\theta)\right.\right.\\ &\left.\left. - \epsilon_1 \epsilon_2 \cos^2\theta \sin^2\theta\right)\right]. \end{aligned} \qquad (S19)$$

For very small deformation, fiber angle distribution stays to be more or less uniform; we can integrate $F_f(\theta)$ over fiber angle to get the total deformation energy as a function of $\epsilon_1, \epsilon_2$:

$$\begin{aligned}F(\epsilon_1, \epsilon_2) &\approx \frac{N_f}{\pi} \int F_f(\theta) d\theta \\ &= \frac{N_f l^2}{4\pi}\left[\left(\frac{5\pi}{8}\epsilon_1^2 + \frac{\pi}{8}\epsilon_2^2 + \frac{\pi}{2}\epsilon_1 \epsilon_2\right)\mu_1 + \left(\frac{\pi}{8}\epsilon_1^2 + \frac{5\pi}{8}\epsilon_2^2 + \frac{\pi}{2}\epsilon_1 \epsilon_2\right)\mu_2\right], \end{aligned} \qquad (S20)$$

with $N_f$ is the total number of fibers. From $F(\epsilon_1, \epsilon_2)$, we can calculate the two principal (linear) elastic moduli:

$$E_1 = \frac{\partial^2 F}{\partial \epsilon_1^2} = \frac{N_f l^2}{16}(5\mu_1 + \mu_2), \qquad (S21a)$$

$$E_2 = \frac{\partial^2 F}{\partial \epsilon_2^2} = \frac{N_f l^2}{16}(\mu_1 + 5\mu_2), \qquad (S21b)$$

from which we obtain

$$\frac{E_2}{E_1} = \frac{(5\mu_1 + \mu_2)}{(\mu_1 + 5\mu_2)}. \qquad (S22)$$

We fitted the simulation results using the following expression

$$\frac{E_2}{E_1} = \frac{(\rho\mu_1 + \mu_2)}{(\mu_1 + \rho\mu_2)}, \qquad (S23)$$

with $\rho \approx 4.1$, which is very close to the prediction $\rho = 5.0$ in Eq. (S22) from affine theory.

## VI. Continuum theory: decay of displacements induced by cell contraction

*1. Linear anisotropic elastic continuum*

In this section, we calculate the decay of displacements induced by a contracting cell in linear anisotropic elastic medium in two dimensions. To be specific, we consider a circularly contracting cell in 2D linear *spherically isotropic* elastic medium, in which there are three independent material parameters, two principal Young's moduli, $E_1$ and $E_2$ (along its radial and tangential axes respectively), and one Poisson's ratio $\nu_{12}$ or $\nu_{21}$ with $\nu_{21}/E_2 = \nu_{12}/E_1$. The corresponding elastic energy density is given by

$$F = \frac{1}{2}\frac{E_1}{1 - \nu_{12}\nu_{21}}\epsilon_1^2 + \frac{E_2\nu_{12}}{1 - \nu_{12}\nu_{21}}\epsilon_1\epsilon_2 + \frac{1}{2}\frac{E_2}{1 - \nu_{12}\nu_{21}}\epsilon_2^2, \qquad (S24)$$

from which we obtain the stress-strain relations as

$$\begin{aligned}\sigma_r &= \frac{E_1}{1 - \nu_{12}\nu_{21}}\epsilon_r + \frac{E_1\nu_{21}}{1 - \nu_{12}\nu_{21}}\epsilon_\theta \\ \sigma_\theta &= \frac{E_2\nu_{12}}{1 - \nu_{12}\nu_{21}}\epsilon_r + \frac{E_2}{1 - \nu_{12}\nu_{21}}\epsilon_\theta\end{aligned}, \qquad (S25)$$

or equivalently,

$$\begin{aligned}\epsilon_r &= \frac{1}{E_1}\sigma_r - \frac{\nu_{21}}{E_2}\sigma_\theta \\ \epsilon_\theta &= -\frac{\nu_{12}}{E_1}\sigma_r + \frac{1}{E_2}\sigma_\theta\end{aligned}. \qquad (S26)$$

For the displacement fields generated by a circularly contracting cell in 2D infinite elastic medium, we have circular symmetry and the two principal strains are given by

$$\epsilon_r = \frac{du}{dR}; \quad \epsilon_\theta = \frac{u}{R}, \quad (S27)$$

where $u$ is the displacement along the radial direction.

The 2D mechanical equilibrium condition is given by

$$\frac{d\sigma_r}{dR} + \frac{\sigma_r - \sigma_\theta}{R} = 0. \quad (S28)$$

Substituting Eq. (S27) into Eqs. (S25) and (S28) subsequently, we obtain the equilibrium equation for displacement field $u(R)$ as

$$\frac{d^2u}{dR^2} + \frac{1}{R}\frac{du}{dR} - \frac{E_2}{E_1}\frac{u}{R^2} = 0. \quad (S29)$$

The general solution of Eq. (S29) is

$$u = aR^n + bR^{-n}, \quad (S30)$$

with

$$n = \sqrt{E_2/E_1}. \quad (S31)$$

Particularly, for cells in an infinite elastic medium with boundary conditions, $\lim_{R\to\infty} u = \text{finite}$ and $u(R = R_{\text{cell}}) = U_{\text{cell}}$, we obtain Eq. (S11) (see Sec. IV).

*2. Strain-induced anisotropic nonlinear elastic continuum*

More generally, if the fibrous network that is contracted by a circular cell can be simply mapped to be a spherically isotropic elastic medium as

$$\begin{aligned}\sigma_r &= \tilde{E}_1\epsilon_r + \tilde{E}_1 v_{21}\epsilon_\theta \\ \sigma_\theta &= \tilde{E}_1 v_{21}\epsilon_r + \tilde{E}_2\epsilon_\theta\end{aligned}, \quad (S32)$$

where $\tilde{E}_1$, $\tilde{E}_2$, and $v_{21}$ are the mapping parameters that usually also depend on strains. Substituting it into Eq. (S28), we can obtain a similar equation as (S29) for the highly nonlinear near-field region, in which we can assume that the radial moduli $\tilde{E}_1$ and Poisson's ratio $v_{21}$ changes very slowly and increases with cell contractions. In this case, the decay of near-field displacement follows an effective power law $\tilde{u} \sim 1/\tilde{R}^n$ with exponent, $n$, determined by relative magnitude between two principal moduli. This is consistent with the simulation data showing a good power-law decay of cell-induced displacements in the near-field region (see Fig. 2B in the main text).

We match the far-field solution and the near-field solution given by

$$\tilde{u}_{\text{far}} = \widetilde{U}_{\text{eff}}/\tilde{R}, \qquad (S33)$$

$$\tilde{u}_{\text{near}} = a\tilde{R}^n + b\tilde{R}^{-n}, \qquad (S34)$$

respectively, using the continuity of the displacement and stress at $R = R_T$. Here $\widetilde{U}_{\text{eff}}$ is the normalized effective far-field displacement, $\widetilde{U}_{\text{eff}}/\tilde{R}_T^2 = \mathcal{A}^{-1}$ with $\mathcal{A} = \mathcal{C}/\mathcal{C}_{\text{cr}}$ being the normalized cell contraction and measuring the nonlinearity of the cell-contracted network; we obtain $a = 1 - b$ and $b = \mathcal{A}^{-1}\frac{n+1}{2n}\tilde{R}_T^{n+1}$ with the matching radius $\tilde{R}_T$ satisfying the equation

$$\frac{1+n}{2n}\tilde{R}_T^{1+n} - \frac{1-n}{2n}\tilde{R}_T^{1-n} - \mathcal{A} = 0. \qquad (S35)$$

The normalized effective far-field displacement given by $\widetilde{U}_{\text{eff}} = \mathcal{A}^{-1}\tilde{R}_T^2$, which can be much larger than that of linear, isotropic medium for which $n = 1$ and $\widetilde{U}_{\text{eff}} = 1$. If $\tilde{R}_T \gg 1$ and $\mathcal{A} \gg 1$, we obtain the linear relation between the effective far-field displacement and the normalized cell contraction, i.e.,

$$\widetilde{U}_{\text{eff}} \sim \frac{\mathcal{C}}{\mathcal{C}_{\text{cr}}}, \qquad (S36)$$

which has been shown to agree with simulation measurements (see Fig. 2D in the main text).

1. Sopher RS, et al. (2018) Nonlinear Elasticity of the ECM Fibers Facilitates Efficient Intercellular Communication. *Biophys J* 115(7):1357–1370.